\documentclass[transmag]{IEEEtran}

\usepackage{multirow}
\usepackage{nicefrac}
\usepackage{etoolbox}
\usepackage{amsmath}
\usepackage{graphicx}
\usepackage{graphics}
\usepackage{subcaption}
\usepackage{textcmds}
\usepackage{hyperref}

\begin{document}

\title{On the Optimization of Underwater Quantum Key Distribution Systems with Time-Gated SPADs}


\author{\IEEEauthorblockN{Amir Hossein Fahim Raouf\IEEEauthorrefmark{1} and Murat Uysal\IEEEauthorrefmark{1}}
\IEEEauthorblockA{\IEEEauthorrefmark{1}Department of Electrical and Electronics Engineering, Ozyegin University, Istanbul, Turkey, 34794.}
\IEEEauthorblockA{amirh.fraouf@ieee.org}
}

\maketitle




\begin{abstract}
In this paper, we study the effect of various transmitter and receiver parameters on the quantum bit error rate (QBER) performance of underwater quantum key distribution. We utilize a Monte Carlo approach to simulate the trajectories of emitted photons transmitting in the water from the transmitter towards receiver. Based on propagation delay results, we first determine a proper value for bit period to avoid the intersymbol interference as a result of possible multiple scattering events. Then, based on the angle of arrival of the received photons, we determine a proper field-of-view to limit the average number of received background noise. Finally, we determine the optimal value for the single photon avalanche diode (SPAD) gate time in the sense of minimizing the QBER for the selected system parameters and given propagation environment.
\end{abstract}

\section{Introduction}\label{sec:intro}
The unconditional security promised by quantum key distribution (QKD) can meet the future data security needs in the post-quantum era \cite{1}. In the last two decades, QKD was successfully demonstrated in fiber optic, atmospheric and satellite links and already commercialized \cite{2}. Another potential application domain of QKD is underwater links with sensitive data such as the submarine communication, surveillance of critical offshore platforms, harbors, ports, etc. \cite{3}. There have been some recent research efforts on the performance analysis and experimental demonstrations of underwater QKD \cite{4, 5, 6, 7, 8, 9, 10, 11, 12, 13, 14}.

In the practical implementation of BB84 protocol \cite{14}, Bob typically employs InGaAs or InP single photon avalanche photodiodes (SPADs) working in the Geiger mode as the detector \cite{15}. One promising approach in such detectors to mitigate the effect of background noise is to utilize gate mode \cite{16}. The gate time, also known as detection time window, is the period during which the detector is active and able to detect a possible coming signal. During the operation, the SPADs are in off-mode most of the time and a narrow gate time is opened only when the signal photons are expected to arrive. Consequently, the noise photons arriving outside the time window can be blocked. The gate however not only eliminates noise, but also potentially eliminates signal counts \cite{17}. Thus, there is a trade-off between the number of background photons and the received signal photons which brings the necessity of finding the optimal value of receiver gate time. For example, Baek \textit{et al.} \cite{18} experimentally investigated the proper selection of the detection time window and the clock frequency for a 50 km fiber link. The optimal selection of gate time obviously depends on the propagation medium. In \cite{19}, Jiang \textit{et al.} studied the optimization problem of gate time period and quantum bit error rate (QBER) performance for the underwater QKD system. A Monte Carlo numerical simulation was utilized to estimate the channel impulse response whish was shown to have a good fit to the double Gamma function model. Based on this channel model, QBER performance for different gate times was reported for different values of environmental irradiance and system configurations for a given receiver field of view (FoV) under the assumption of constant bit period.

In this paper, we investigate the effect of various transmitter and receiver parameters (beam width, beam divergence, bit period, SPAD gate time and FoV) on the performance of underwater QKD. Based on extensive Monte Carlo simulation results, we first determine the propagation delay for a given link distance and accordingly determine a proper value for bit period to avoid the intersymbol interference that might occur due to multiple scattering. Then, based on the angle of arrival (AoA) of the received photons, we determine a proper FoV to reduce the average number of received background noise. For the selected system parameters and given propagation environment, we finally determine the optimal value for the SPAD gate time in the sense of minimizing the QBER.

The rest of the paper is organized as follows: In Section \ref{sec:sys_mod}, we present the system model under consideration. In Section \ref{sec:ch_mod}, we describe the channel modelling approach adopted in our work. In Section \ref{sec:sim}, we present simulation results and discuss the optimal selection of relevant parameters. We finally conclude in Section \ref{sec:conc}.

\section{System Model}\label{sec:sys_mod}
We consider a QKD system employing BB84 protocol for key distribution over a link distance of $L$. This protocol creates a secret key between the authorized partners, Alice and Bob, such that eavesdropper (Eve) cannot acquire noteworthy information. Alice prepares a qubit by choosing randomly between two orthogonal bases, i.e., ${0^{\circ} }/ + {90^{\circ} }$ basis and $ + {45^{\circ} }/ - {45^{\circ} }$ basis for every bit she wants to send. She selects a random bit value ``0'' or ``1'' and uses polarization encoding of photons where polarization of ${0^{\circ} }/ - {45^{\circ} }$ represents 0 and polarization of $ + {90^{\circ} }/ + {45^{\circ} }$ represents 1. After travelling over the underwater channel, the received qubit is randomly fed to two detectors (corresponding to two different bases) by the beam splitter where half wave plate converts the states $ - {45^{\circ} }$ and $ + {45^{\circ} }$ to states ${0^{\circ} }$ and $ + {90^{\circ} }$, and a polarizing splitter redirects it to the corresponding decoder according to the polarization state of the received photon. The secure key between two parties is constructed based on the ``sift'' events which correspond to the bit intervals in which exactly one of the SPADs registers a count and both Alice and Bob have chosen the same basis.

\begin{figure}[tb]
\centering
\includegraphics[trim=0.3cm 0.65cm 0.5cm 0.1cm,width=\linewidth]{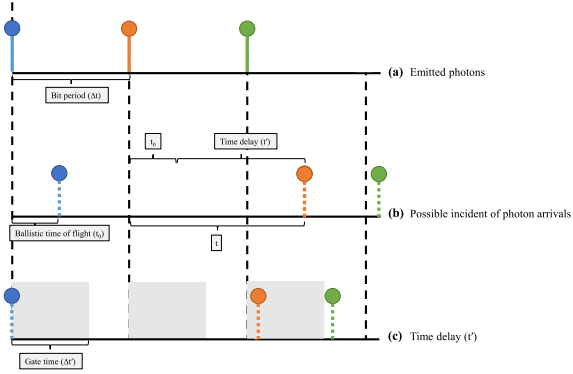}
\caption{An example of emitted photons and possible incident photon arrivals in the present of SPAD gating functionality.}
\label{fig1}
\end{figure}

In a homogeneous medium, the required time for a photon to travel a path length of $L'$ can be calculated as $t = {{L'} \mathord{\left/
 {\vphantom {{L'} {{c_n}}}} \right.
 \kern-\nulldelimiterspace} {{c_n}}} = {{L'n} \mathord{\left/
 {\vphantom {{L'n} {{c_0}}}} \right.
 \kern-\nulldelimiterspace} {{c_0}}}$ where ${c_n}$ is the speed of light in a medium with index of refraction $n$ \cite{20}. As an example, Fig. \ref{fig1}(a) illustrates three consecutive bit periods (denoted by $\Delta t$) in which a different colored photon is emitted. Fig. \ref{fig1}(b) shows a realization of the possible incident photon arrivals. Here, ${t_0}$ is the ballistic time of flight \cite{21} and represents the minimum travel time, i.e., assuming a straight line distance and no scattering) between the receiver and transmitter. In this example, the blue colored photon does not experience any scattering and arrives at ${t_0}$. On the other hand, orange and green colored photons experience additional time delay. Let $t'$ denote the time delay and represent the additional time that the photon takes to arrive at the receiver due to scattering in the underwater environment. This can be calculated as $t' = t - {t_0}$. Fig. \ref{fig1}(c) depicts the time delay of incident photon arrivals by removing the ballistic time of flight (i.e., ${t_0}$) where the shaded area represents the gate time window (i.e., $\Delta t'$). In this example, the time delay of orange colored photon is larger than the bit period (i.e., $\Delta t < t'$) which leads to an intersymbol interference. The time delay of green colored photon is less than the bit period but is larger than the gate time (i.e., $\Delta t' < t' < \Delta t$) which results in missing the signal photon at the receiver side.

Let ${n_S}$ denote the average photon numbers per bit period. Under the assumption of ideal single photon transmitter (i.e., one single photon per pulse), we can simply set ${n_S} = 1$. Due to the inevitable photon interactions with water molecules and other particles in water, the emitted photons experience absorption and scattering. As illustrated in Fig. \ref{fig1}, multiple scattering events might result in intersymbol interference, i.e., an emitted photon within the previous bit period can arrive late due to scattering and therefore recorded in the next period

To prevent symbol interference, we need to set the minimum bit period $\Delta t$ as the required average time for receiving a certain percentage (e.g., 99.9\%) of total received photons over different trials\footnote{In our simulation, we launch a photon in each trial and trace it until it reaches the receiver plane. This is repeated multiple times.}. In addition to the transmitted photons, the receiver will collect ${n_B}$ background photons on average. This is given by
   
\begin{equation}
    {n_B} = {{\pi {R_d}A\lambda \Delta \lambda \left( {1 - \cos \left( \Omega  \right)} \right)\Delta t'} \mathord{\left/
 {\vphantom {{\pi {R_d}A\lambda \Delta \lambda \left( {1 - \cos \left( \Omega  \right)} \right)\Delta t'} {\left( {2{h_p}{c_0}} \right)}}} \right.
 \kern-\nulldelimiterspace} {\left( {2{h_p}{c_0}} \right)}}
\end{equation}
where $\Delta t'$ is the receiver gate time, ${R_d}$ is the irradiance of the environment whose primary source is the refracted sunlight from the surface of the water \cite{22}, $A$ is the receiver aperture area, $\lambda $ is the wavelength, $\Delta \lambda$ is the filter spectral width, $\Omega$ is the field of view, ${h_p}$ is Planck's constant, and ${c_0}$ is the speed of light. In addition to the background noise, the receiver will be subject to an average equivalent dark current count (denoted as ${I_{dc}}$) which induces an ${n_D} = {I_{dc}}\Delta t$ background photons. Therefore, the average number of noise photons reaching each Bob's detector can be expressed as \cite{23}

\begin{equation}
{n_N} = {n_D} + {{{n_B}} \mathord{\left/
 {\vphantom {{{n_B}} 2}} \right.
 \kern-\nulldelimiterspace} 2}
\end{equation}

As mentioned above, underwater QKD communication is subject to absorption and scattering. Let $\gamma$ denote the fraction of received photons \cite{23}. To ensure correct decoding of the received qubits, each Alice and Bob need to sift the bits that are encoded and decoded using the same bases. Sifting process will result in 50\% of correctness due to the deviant selection on bases at both Alice and Bob \cite{24}. QBER is defined as the error rate in the sifted key and can be calculated as \cite{23}

\begin{equation}\label{eq:3}
\text{QBER} = \frac{{{n_N}}}{{{{\gamma {n_S}} \mathord{\left/
 {\vphantom {{\gamma {n_S}} 2}} \right.
 \kern-\nulldelimiterspace} 2} + 2{n_N}}}
\end{equation}
where the factor of $\nicefrac{1}{2}$ is introduced as a result of sifting process.

\section{Channel Modelling Approach}\label{sec:ch_mod}
\begin{figure}[tb]
\centering
\includegraphics[width=0.8\linewidth]{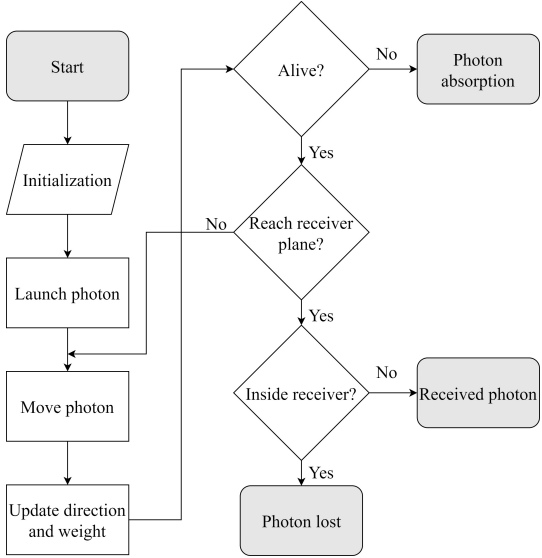}
\caption{The flowchart of Monte Carlo simulation of light propagation.}
\label{fig2}
\end{figure}

In this work, we adopt Monte Carlo modeling of light transport in multi-layered tissues (MCML) method \cite{25} for channel modeling. The major steps in the simulation are summarized in Fig. \ref{fig2}. The transmitter is represented as a circle with an initial radius of ${r_0}$ and maximum initial divergence angle of ${\theta _{0,\max}}$ located in $z = 0$ plane. Let $U\left[ {n,m} \right]$ denote the uniform random variable (RV) with a distribution between $n$ and $m$. The initial location of each photon $\left( {x,y,0} \right)$ is specified by $x = {r_0}\cos {\varphi _0}$ and $y = {r_0}\sin {\varphi _0}$ where the azimuthal angle ${\varphi _0}$ is generated based on $U\left[ {0,2\pi } \right]$. The direction $\left( {{\mu _x},{\mu _y},{\mu _z}} \right)$ of the photon can be expressed as ${\mu _x} = \sin {\theta _0}\cos {\varphi _0}$, ${\mu _y} = \sin {\theta _0}\sin {\varphi _0}$ and ${\mu _z} = \cos {\theta _0}$ where the polar angle ${\theta _0}$ is generated according to $U[ - {\theta _{0,\max }},{\theta _{0,\max }}]$. 

The initial weight of the launched photon is considered as 1. The emitted photon travels a random distance before interacting with a particle in the medium. In homogenous water, the geometric path length $\delta $ can be computed with $\delta = - {{\ln \left( q \right)} \mathord{\left/
 {\vphantom {{\ln \left( q \right)} {\varsigma \left( \lambda \right)}}} \right.
 \kern-\nulldelimiterspace} {\varsigma \left( \lambda  \right)}}$ where $q$ is $U\left[ {0,1} \right]$ and $\varsigma \left( \lambda  \right) = \alpha \left( \lambda  \right) + \beta \left( \lambda  \right)$ is the extinction coefficient where $\alpha \left( \lambda  \right)$ and $\beta \left( \lambda  \right)$ respectively denote absorption and scattering coefficients. We assume that the quantum states of the scattered photons remain invariant \mbox{\cite{7, 11,12,13,26new}}.

 The photon loses a fraction of its initial weight and deviates from its starting path as a result of interaction with the particles. Let $w$ and $w'$ denote the photon weight before and after the interaction, respectively. The photon weight can be updated as $w' = w{{\beta \left( \lambda  \right)} \mathord{\left/
 {\vphantom {{\beta \left( \lambda  \right)} {\varsigma \left( \lambda  \right)}}} \right.
 \kern-\nulldelimiterspace} {\varsigma \left( \lambda  \right)}}$.

Let $\left( {{{\mu '}_x},{{\mu '}_y},{{\mu '}_z}} \right)$ denote the new propagation direction. For a given azimuthal angle $\phi$ and scattering angle $\theta$, they can be defined respectively as 

\begin{equation}
{\mu '_x} = {{\left( {{\mu _x}{\mu _z}\cos \phi  - {\mu _y}\sin \phi } \right)\sin \theta } \mathord{\left/
 {\vphantom {{\left( {{\mu _x}{\mu _z}\cos \phi  - {\mu _y}\sin \phi } \right)\sin \theta } {\sqrt {1 - \mu _z^2} }}} \right.
 \kern-\nulldelimiterspace} {\sqrt {1 - \mu _z^2} }} + {\mu _x}\cos \theta 
\end{equation}

\begin{equation}
{\mu '_y} = {{\left( {{\mu _y}{\mu _z}\cos \phi  + {\mu _x}\sin \phi } \right)\sin \theta } \mathord{\left/
 {\vphantom {{\left( {{\mu _y}{\mu _z}\cos \phi  + {\mu _x}\sin \phi } \right)\sin \theta } {\sqrt {1 - \mu _z^2} }}} \right.
 \kern-\nulldelimiterspace} {\sqrt {1 - \mu _z^2} }} + {\mu _y}\cos \theta 
\end{equation}

\begin{equation}
{\mu '_z} =  - \sin \theta \cos \phi \sqrt {1 - \mu _z^2}  + {\mu _z}\cos \theta 
\end{equation}

The new position of the photon can be therefore updated by $x' = x + {\mu '_x}\delta$, $y' = y + {\mu '_y}\delta$ and $z' = z + {\mu '_z}\delta$. If the photon direction is close to z-axes (i.e., $\left| {{\mu _z}} \right| > 0.9999$), then the new propagation direction can be approximated as ${\mu '_x} \cong \sin \theta \cos \phi$, ${\mu '_y} \cong \sin \theta \cos \phi$ and ${\mu '_z} \cong {{{\mu _z}\cos \phi } \mathord{\left/
 {\vphantom {{{\mu _z}\cos \phi } {\left| {{\mu _z}} \right|}}} \right.
 \kern-\nulldelimiterspace} {\left| {{\mu _z}} \right|}}$.

The azimuthal angle $\phi$ is modeled as an RV with a distribution of $U[0,2\pi ]$. The scattering angle $\theta$ can be calculated based on the scattering phase function which provides the angular distribution of light intensity scattered by a particle \cite{25}. The channel characteristics for collimated beams highly depend on the scattering phase function \cite{25}. An RV $\chi$ is generated randomly with a distribution of $U\left[ {0,\pi } \right]$, and then the corresponding $\theta$ can be calculated as 

\begin{equation}
\chi  = \int_0^\theta  {p\left( {\Psi ,a,{g_{\rm{f}}},{g_{\rm{B}}}} \right)\sin \left( \Psi  \right)d\Psi } 
\end{equation}
where $p\left( {\Psi ,a,{g_{\rm{f}}},{g_{\rm{B}}}} \right)$ denotes scattering phase function. In our work, we utilize the two term Henyey–Greenstein (TTHG) model as the scattering phase function which is shown to have a good fit with the experimental results \cite{25}. It is defined as

\begin{equation}
{p_{{\rm{TTHG}}}}\left( {\theta ,a,{g_{\rm{f}}},{g_{\rm{B}}}} \right) = a{p_{{\rm{HG}}}}\left( {\theta ,{g_{\rm{F}}}} \right) + \left( {1 - a} \right){p_{{\rm{HG}}}}\left( {\theta , - {g_{\rm{B}}}} \right)
\end{equation}
where $a$ is the weight of the forward Henyey–Greenstein (HG) phase function, and ${g_{\rm{F}}}$ and ${g_{\rm{B}}}$ are the asymmetry factors for the forward and backward-directed HG phase functions, respectively \cite{26}. ${p_{{\rm{HG}}}}\left( {\theta ,g} \right)$ is HG function and is given by 

\begin{equation}
{p_{{\rm{HG}}}}\left( {\theta ,g} \right) = {{\left( {1 - {g^2}} \right)} \mathord{\left/
 {\vphantom {{\left( {1 - {g^2}} \right)} {\left[ {2{{\left( {1 + {g^2} - 2g\cos \theta } \right)}^{1.5}}} \right]}}} \right.
 \kern-\nulldelimiterspace} {\left[ {2{{\left( {1 + {g^2} - 2g\cos \theta } \right)}^{1.5}}} \right]}}
\end{equation}

The relationships between $a$, ${g_{\rm{B}}}$, ${g_{\rm{F}}}$, and $\overline {\cos \theta }$ can be expressed as \cite{26} 

\begin{equation}
{g_{\rm{B}}} =  - 0.3061446 + 1.000568{g_{\rm{F}}} - 0.01826338g_{\rm{F}}^2 + 0.03643748g_{\rm{F}}^3
\end{equation}
\begin{equation}
a = {{{g_{\rm{B}}}\left( {1 + {g_{\rm{B}}}} \right)} \mathord{\left/
 {\vphantom {{{g_{\rm{B}}}\left( {1 + {g_{\rm{B}}}} \right)} {\left[ {\left( {{g_{\rm{F}}} + {g_{\rm{B}}}} \right)\left( {1 + {g_{\rm{B}}} - {g_{\rm{F}}}} \right)} \right]}}} \right.
 \kern-\nulldelimiterspace} {\left[ {\left( {{g_{\rm{F}}} + {g_{\rm{B}}}} \right)\left( {1 + {g_{\rm{B}}} - {g_{\rm{F}}}} \right)} \right]}}
\end{equation}
\begin{equation}
\overline {\cos \theta }  = a\left( {{g_{\rm{F}}} + {g_{\rm{B}}}} \right) - {g_{\rm{B}}}
\end{equation}

An approximation for $\overline {\cos \theta }$ is presented in \cite{26} as $\overline {\cos \theta }  \cong {{2\left( {1 - 2B} \right)} \mathord{\left/
 {\vphantom {{2\left( {1 - 2B} \right)} {\left( {2 + B} \right)}}} \right.
 \kern-\nulldelimiterspace} {\left( {2 + B} \right)}}$ where $B = {{{\beta _b}} \mathord{\left/
 {\vphantom {{{\beta _b}} \beta }} \right.
 \kern-\nulldelimiterspace} \beta }$ and ${\beta _b}$ is back scattering coefficient.

The cycle of traveling and interacting with particles proceeds until, whether the photon is absorbed or it reaches the receiver plane. The photon is assumed to be absorbed when its weight becomes less than a threshold denoted as ${w_{th}}$. The photon is recorded as received if its location is within the radius of the aperture (denoted as $d$) and its azimuthal AoA\footnote{The AoA is defined as the incident angle with respect to the optical axis and the zero AoA represents that the direction of received photon is parallel to the optical axis.} is less than FoV of the receiver. In addition, the time delay of a received photon (denoted by $t'$) should be less than the receiver gate time; i.e., $t' \le \Delta t'$. For example in Fig. \ref{fig1}(c), the time delay of the third emitted photon (green color) is longer than the receiver gate time; thus the SPAD is unable to detect this photon.

\begin{table}[tb]
\captionsetup{justification=centering}
\caption{System and channel parameters}
\label{table1}
\begin{center}
\scalebox{0.8}{
\begin{tabular}{ |l|l|l| } 
 \hline
 \textbf{Parameter} & \textbf{Definition} & \textbf{Numerical Value} \\ \hline
 ${r_0}$ & \text{Transmitter beam width} & 3 mm \cite{25} \\ \hline 
${\theta _{0,\max }}$ & \text{Maximum beam divergence} & ${20^{\circ} }$ \cite{25} \\ \hline 
 $\Delta \lambda$ & \text{Filter spectral width} & $30$ $\rm{nm}$ \cite{22} \\ \hline 
$\lambda$ & \text{Wavelength} & $532$ $\rm{nm}$ \cite{22} \\ \hline 
$d$ & \text{Receiver aperture diameter} & $20$ $\rm{cm}$ \cite{25} \\ \hline 
$I_{dc}$ & \text{Dark current count rate} & $60$ $\rm{Hz}$ \cite{22} \\ \hline 
$K_{\infty}$ & \text{Asymptotic diffuse attenuation coefficient} & $0.08$ $\rm{m^{-1}}$ \cite{28} \\ \hline 
$z_d$ & \text{Depth} & $100$ $\rm{m}$ \cite{22} \\ \hline 
${R_d}\left( {\lambda ,0} \right)$ & \text{Irradiance of the underwater environment at sea surface} & ${10^{ - 3}}\,{{\rm{W}} \mathord{\left/
 {\vphantom {{\rm{W}} {{{\rm{m}}^{\rm{2}}}}}} \right.
 \kern-\nulldelimiterspace} {{{\rm{m}}^{\rm{2}}}}}$ \cite{27} \\ \hline $\varsigma $ & \text{Extinction coefficient (clear water)} & 0.151 m$^{-1}$ \cite{25} \\
\hline
\end{tabular}}
\end{center}
\end{table}

\section{Simulation Results and Discussion}\label{sec:sim}
 
In this section, we present simulation results and discuss the optimal selection of transmitter and receiver parameters. We first investigate the time delay and AoA for choosing the proper values for bit period and receiver FoV, respectively. Using these values, we identify the optimal value of the gate time in the sense of minimizing the QBER performance. In our simulation study, we assume a transmitter beam width of ${r_0} = $ 3 mm \cite{25}, a maximum beam divergence of ${\theta _{0,\max }} = {20^{\circ} }$ \cite{25}, the dark current count rate of ${I_{dc}} = 60$ Hz \cite{22}, filter spectral width of $\Delta \lambda = 30$ nm \cite{22}, the wavelength of $\lambda = $ 532 nm \cite{25}, the receiver aperture diameter of $d = 20$ cm \cite{25}, an extinction coefficient of $\varsigma = 0.151\,{{\rm{m}}^{ - 1}}$ \cite{25} (corresponding to clear ocean conditions), $\overline {\cos \theta } = 0.9675$ \cite{25}, a depth of ${z_d} = $ 100 m \cite{22}, the asymptotic diffuse attenuation coefficient of ${K_\infty } = 0.08$ \cite{28} and clear atmospheric conditions at night with a full moon, i.e., ${R_d}\left( {\lambda ,0} \right) = {10^{ - 3}}\,{{\rm{W}} \mathord{\left/
 {\vphantom {{\rm{W}} {{{\rm{m}}^{\rm{2}}}}}} \right.
 \kern-\nulldelimiterspace} {{{\rm{m}}^{\rm{2}}}}}$ \cite{27}. For the convenience of the reader, the channel and system parameters are summarized in Table \ref{table1}. In our simulations, we have generated $10^8$ photons for each experiment to obtain reliable results.

\begin{figure}[tb]
\centering
\includegraphics[trim=0.3cm 0.65cm 0.5cm 0.1cm,width=0.6\linewidth]{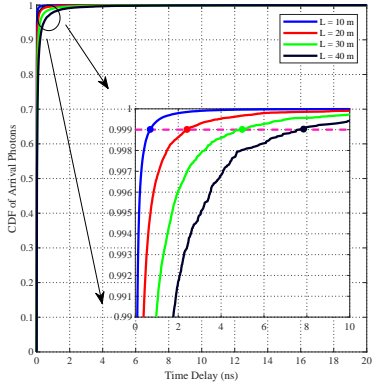}
\caption{CDF of arrival photons based on time of arrival for clear ocean and link distance of 10, 20, 30 and 40 m.}
\label{fig3}
\end{figure}

First, we investigate the time delay for the received photons over various link distances to determine the bit period. In our simulations, we generate $10^8$ photons that travel from the transmitter to the receiver through the underwater channel. At the receiver side, we record the time delay for each received photon. Fig. \ref{fig3} illustrates the cumulative distribution function (CDF) of received photons with respect to the time delay for $L = $ 10, 20, 30 and 40 m. We assume an FoV of ${180^{\circ} }$ to consider the worst case concerning symbol interference. As it can be observed from Fig. \ref{fig3}, the time delay of received photons increases as the link distance increases due to the increasing possibility of multiple scattering events. Thus, the bit period depends on the link distance. The required time to receive 99.9\% of the total received photons for link distance of $L = $ 10, 20, 30 and 40 m are respectively $\Delta t$ = 1, 3, 5, and 8 ns. These values are considered as the bit period for the corresponding underwater QKD system under consideration. 

\begin{figure}[tb]
\centering
\includegraphics[trim=0.3cm 0.65cm 0.5cm 0.1cm,width=0.6\linewidth]{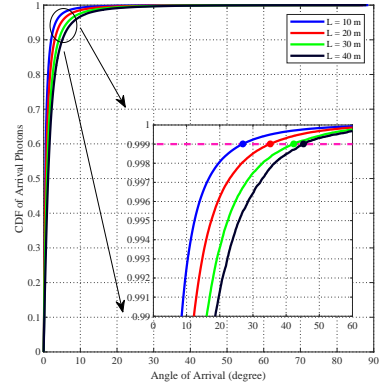}
\caption{CDF of arrival photons based on angle of arrival for clear ocean and link distance of 10, 20, 30 and 40 m.}
\label{fig4}
\end{figure}

Next, we study the AoA for the received photons to determine a proper FoV which can affect the number of received signal photons as well as the average number of background photons. Similar to the previous figure, $10^8$ photons are generated in the simulations and the AoA for each received photon is recorded. Fig. \ref{fig4} illustrates the CDF of received photons versus AoA. As it can be observed from Fig. 4, the required FoV to receive 99.9\% of the total received photons for link distance of $L$ = 10, 20, 30 and 40 m are $\Omega = {27^{\circ} }$, ${35^{\circ} }$, ${42^{\circ} }$, and ${45^{\circ} }$, respectively.

\begin{figure}
     \centering
     \begin{subfigure}[tb]{0.4\linewidth}
         \centering
         \captionsetup{justification=centering}
         \includegraphics[trim=0.3cm 0.7cm 0.5cm 0.5cm,width=\textwidth]{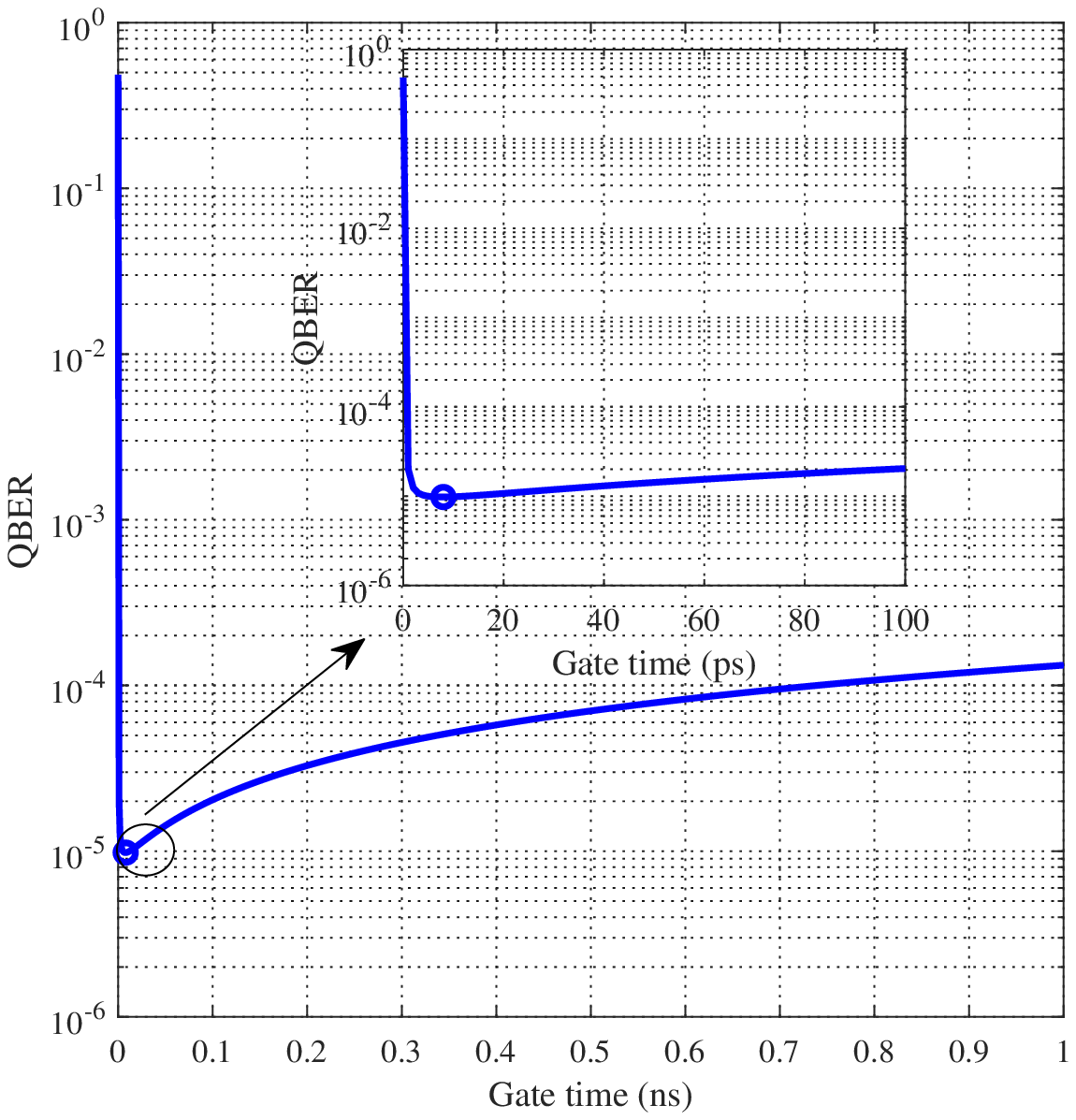}
         \caption{{}}
         \label{fig5a}
     \end{subfigure}
     \begin{subfigure}[tb]{0.4\linewidth}
         \centering
         \captionsetup{justification=centering}
         \includegraphics[trim=0.3cm 0.65cm 0.5cm 0.1cm,width=\textwidth]{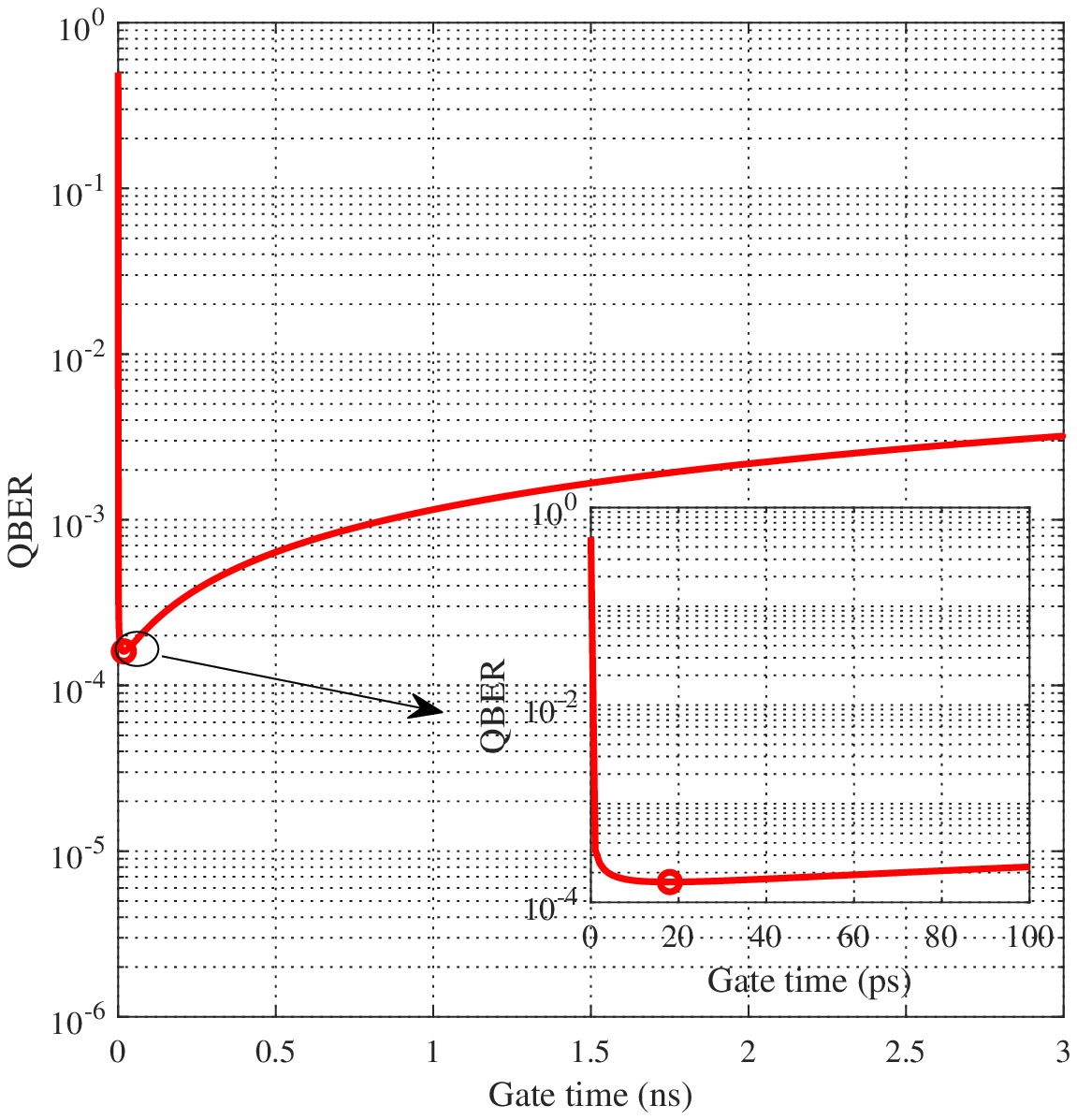}
         \caption{{}}
         \label{fig5b}
     \end{subfigure}
     \par
       \begin{subfigure}[tb]{0.4\linewidth}
         \centering
         \captionsetup{justification=centering}
         \includegraphics[trim=0.3cm 0.7cm 0.5cm 0.5cm,width=\textwidth]{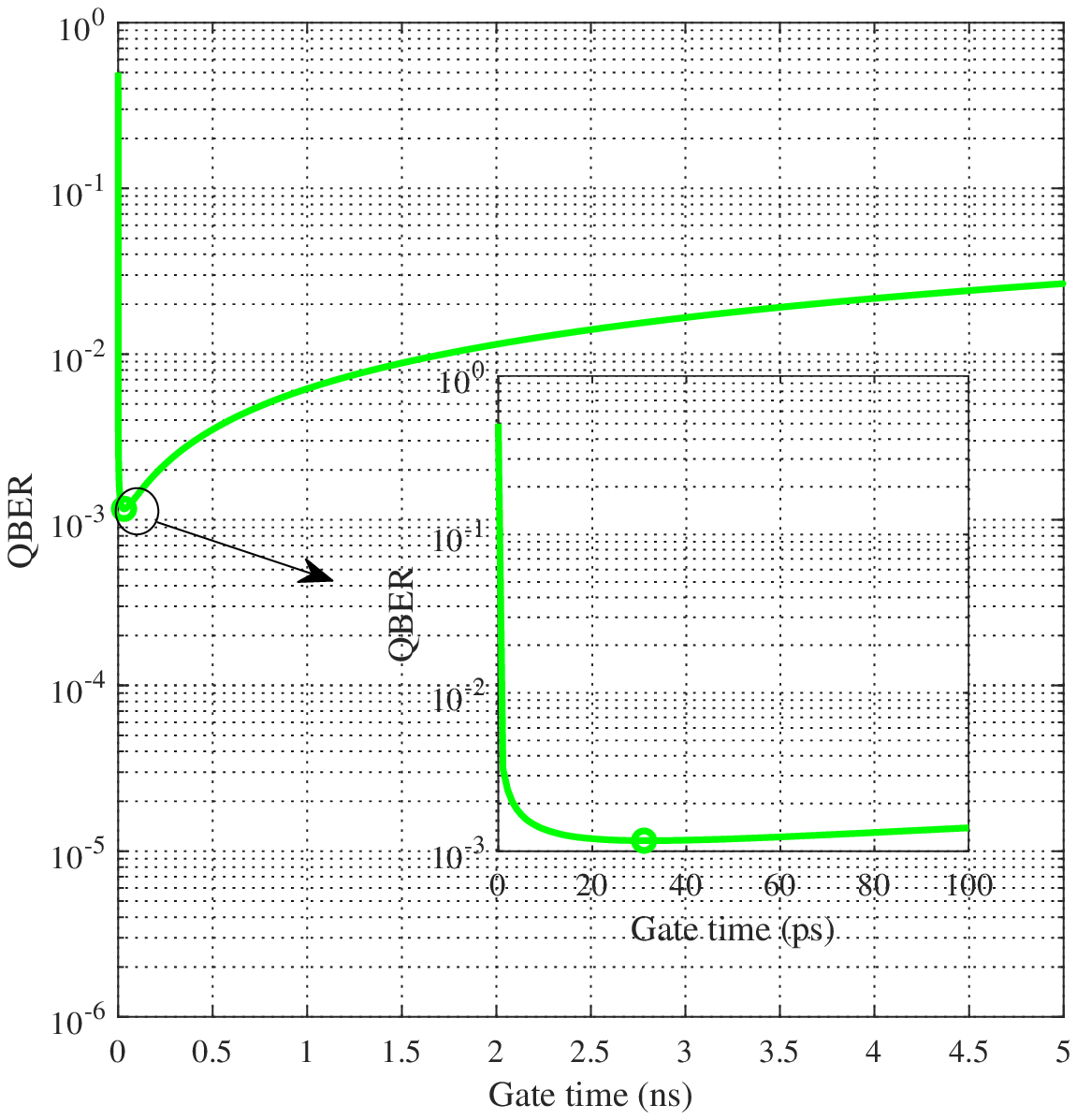}
         \caption{{}}
         \label{fig5c}
     \end{subfigure}
     \begin{subfigure}[tb]{0.4\linewidth}
         \centering
         \captionsetup{justification=centering}
         \includegraphics[trim=0.3cm 0.65cm 0.5cm 0.1cm,width=\textwidth]{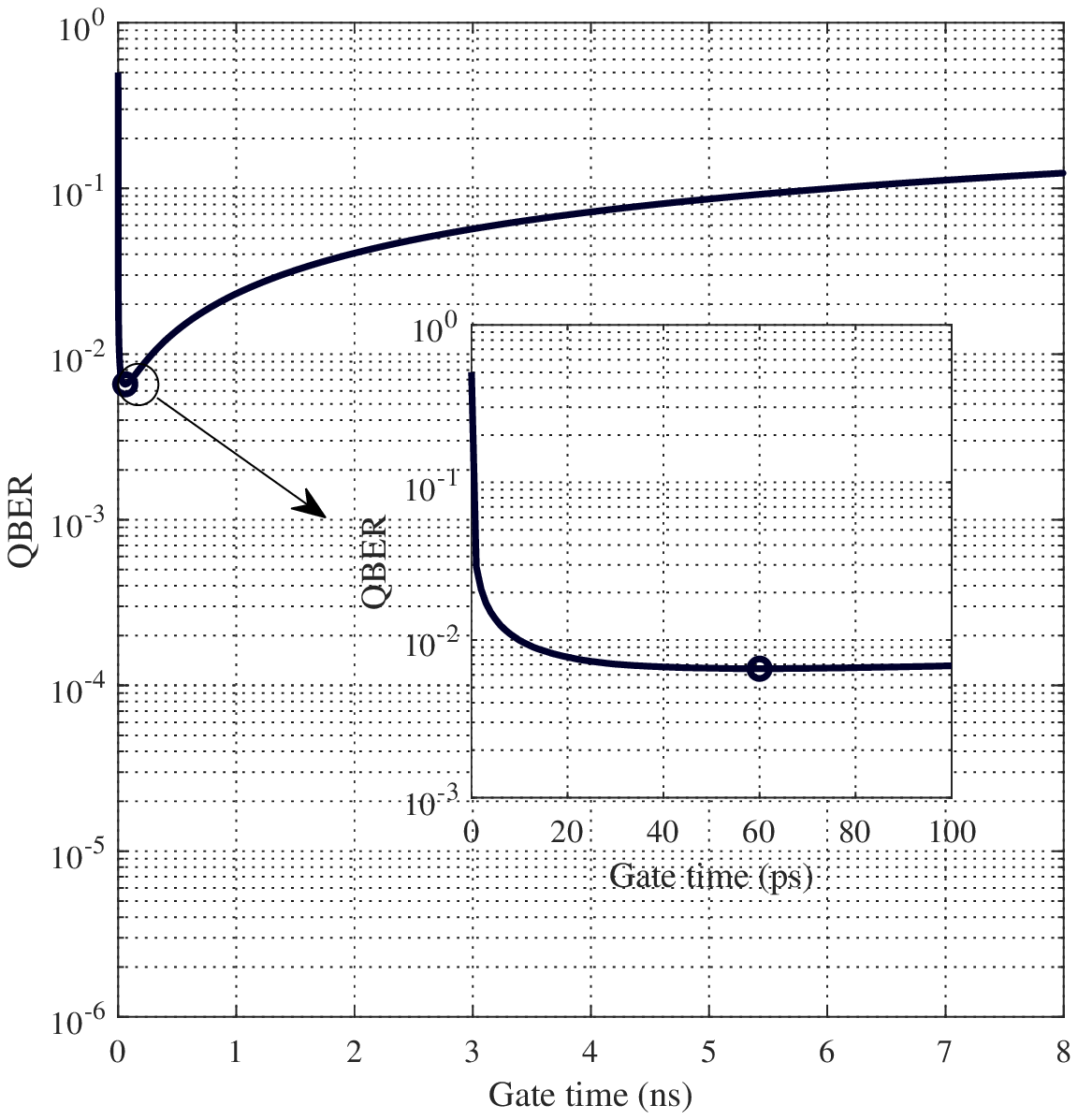}
         \caption{{}}
         \label{fig5d}
     \end{subfigure}
    \caption{QBER performance with respect to the gate time for clear ocean and link distance of \textbf{a)} $L$ = 10 m, \textbf{b)} $L$ = 20 m, \textbf{c)} $L$ = 30 m and \textbf{d)} $L$ = 40 m}
        \label{fig5}
\end{figure}

After we select the values of bit period and FoV for a given link distance, we present the effect of gate time on the QBER performance for underwater QKD system in Fig. \ref{fig5}. The QBER performance is plotted based on the expression in \eqref{eq:3}. The performance curves indicate that there is trade-off between the average number of background photons and the received photons. Through exhaustive search, we find the optimal value of gate time (in the sense of minimizing QBER) as $\Delta t' = $ 9, 19, 32 and 61 ps for link distance of 10, 20, 30 and 40 m, respectively. Although longer gate time increases the average number of received photons, it also increases the background noise counts which leads to a degradation in QBER performance.

Recall that ${r_0} = $ 0.3 cm and ${\theta _{0,\max }} = {20^{\circ} }$ were assumed in Figs. \ref{fig3} and \ref{fig4}. In the following, we further discuss the effect of transmitter beam width (${r_0}$) and maximum beam divergence (${\theta _{0,\max }}$). The resulting optimized values of bit period, FoV and gate time for different combinations for ${r_0}$ and ${\theta _{0,\max }}$ are presented in Table \ref{table2}. First for a fixed value of ${\theta _{0,\max }} = {20^{\circ} }$, we assume ${r_0} = $ 0.3, 3 and 30 cm. We observe that as the beam width increases where the photons are being spread over a larger surface area, the optimal value of gate time increases for the link distance under consideration. For example, the optimal value of the gate time for ${r_0} = $ 0.3, 3 and 30 cm at $L$ = 40 m are respectively $\Delta t' = $ 61, 66, and 107 ps.

Then, for a fixed value of ${r_0} = $ 0.3 cm, we assume ${\theta _{0,\max }} = 0^{\circ}$ and $45^{\circ}$ where ${\theta _{0,\max }} = 0^{\circ}$ indicates the hypothetical case of collimated source. We observe that as the beam divergence increases where the photons are more deviated from the optical axis, the optimal value of gate time increases. For example, the optimal gate time for ${\theta _{0,\max }} = 0^{\circ}$, $20^{\circ}$, and $45^{\circ}$ for link distance of $L$ = 40 m are respectively $\Delta t' = $ 25, 61, and 66 ps.

\begin{table}[tb]
\captionsetup{justification=centering}
\caption{Bit period, FoV, and optimal value of gate time for the given set of system parameters.}
\label{table2}
\begin{center}
\scalebox{0.8}{
\begin{tabular}{|l|l|l|l|l|l|l|}
\hline
${r_0}$ (cm)         & ${\theta _{0,\max }}$ & $L$ (m) & $\Delta t$ (ns) & $\Omega $ & $\Delta t'$ (ps) & QBER      \\ \hline
\multirow{4}{*}{0.3} & \multirow{4}{*}{20$^{\circ}$ }  
     & 10    & 1  & 27$^{\circ}$   & 9   & $9.80\times10^{-6}$ \\ \cline{3-7} 
&    & 20    & 3  & 35$^{\circ}$   & 19   & $1.60\times10^{-4}$ \\ \cline{3-7} 
&    & 30    & 5  & 42$^{\circ}$   & 32   & $1.20\times10^{-3}$ \\ \cline{3-7} 
&    & 40    & 8  & 45$^{\circ}$   & 61   & $6.60\times10^{-3}$ \\ \hline
\multirow{4}{*}{\textbf{3}}   & \multirow{4}{*}{20$^{\circ}$ }  
     & 10    & 1  & 30$^{\circ}$   & 8    & $1.34\times10^{-5}$ \\ \cline{3-7} 
&    & 20    & 3  & 35$^{\circ}$   & 20   & $2.00\times10^{-4}$ \\ \cline{3-7} 
&    & 30    & 6  & 44$^{\circ}$   & 38   & $1.60\times10^{-3}$ \\ \cline{3-7} 
&    & 40    & 11 & 47$^{\circ}$   & 66   & $9.20\times10^{-3}$ \\ \hline
\multirow{4}{*}{\textbf{30}}  & \multirow{4}{*}{20$^{\circ}$ }  
     & 10    & 5  & 58$^{\circ}$   & 70   & $2.80\times10^{-3}$ \\ \cline{3-7} 
&    & 20    & 6  & 55$^{\circ}$   & 63   & $3.10\times10^{-3}$ \\ \cline{3-7} 
&    & 30    & 10 & 53$^{\circ}$   & 81   & $8.20\times10^{-3}$ \\ \cline{3-7} 
&    & 40    & 12 & 51$^{\circ}$   & 107  & $2.10\times10^{-2}$ \\ \hline
\multirow{4}{*}{0.3} & \multirow{4}{*}{\textbf{0$^{\circ}$} }   
     & 10   & 0.06 & 18$^{\circ}$  & 3    & $8.42\times10^{-7}$ \\ \cline{3-7} 
&    & 20   & 0.16 & 20$^{\circ}$  & 6    & $1.21\times10^{-5}$ \\ \cline{3-7} 
&    & 30   & 0.38 & 23$^{\circ}$  & 14   & $1.25\times10^{-4}$ \\ \cline{3-7} 
&    & 40   & 0.73 & 26$^{\circ}$  & 25   & $8.84\times10^{-4}$ \\ \hline
\multirow{4}{*}{0.3} & \multirow{4}{*}{\textbf{45$^{\circ}$} }  
    & 10    & 1   & 27$^{\circ}$   & 8     & $9.78\times10^{-6}$ \\ \cline{3-7} 
&   & 20    & 3   & 37$^{\circ}$   & 17    & $1.63\times10^{-4}$ \\ \cline{3-7} 
&   & 30    & 7   & 44$^{\circ}$   & 39    & $1.50\times10^{-3}$  \\ \cline{3-7} 
&   & 40    & 11  & 48$^{\circ}$   & 66    & $9.10\times10^{-3}$  \\ \hline
\end{tabular}}
\end{center}
\end{table}

\section{Conclusions }\label{sec:conc}
In this paper, we have investigated the effect of various system parameters (beam width, beam divergence, bit period, SPAD gate time and FoV) on the performance of underwater QKD. Based on the propagation delay for a given link distance, the bit period was determined to avoid the possible intersymbol interference due to the multiple scattering. Based on the AoA of the received photons, a proper FoV was selected to minimize the effect of background noise. While longer gate time increases the average number of received photons, it also increases the background noise counts which leads to a degradation in QBER performance. To address this, we have optimized the SPAD gate time in the sense of minimizing the QBER for the selected system parameters and given propagation environment. It has been observed that as the beam width increases where the photons are being spread over a larger surface area, the optimal value of gate time increases for a given link distance. It has been also observed that increasing the divergence angle leads to an increase in the optimal value of gate time.

\vspace{0.5cm}
\noindent\textbf{Disclosures.} The authors declare no conflicts of interest.

\noindent\textbf{Data availability.} No data were generated or analyzed in the presented research.

\bibliographystyle{IEEEtran}
\bibliography{Amir_Gatetime}

\begin{thebibliography}{10}
\providecommand{\url}[1]{#1}
\csname url@samestyle\endcsname
\providecommand{\newblock}{\relax}
\providecommand{\bibinfo}[2]{#2}
\providecommand{\BIBentrySTDinterwordspacing}{\spaceskip=0pt\relax}
\providecommand{\BIBentryALTinterwordstretchfactor}{4}
\providecommand{\BIBentryALTinterwordspacing}{\spaceskip=\fontdimen2\font plus
\BIBentryALTinterwordstretchfactor\fontdimen3\font minus
  \fontdimen4\font\relax}
\providecommand{\BIBforeignlanguage}[2]{{%
\expandafter\ifx\csname l@#1\endcsname\relax
\typeout{** WARNING: IEEEtran.bst: No hyphenation pattern has been}%
\typeout{** loaded for the language `#1'. Using the pattern for}%
\typeout{** the default language instead.}%
\else
\language=\csname l@#1\endcsname
\fi
#2}}
\providecommand{\BIBdecl}{\relax}
\BIBdecl

\bibitem{1}
L.~O. Mailloux, C.~D. Lewis~II, C.~Riggs, and M.~R. Grimaila, ``Post-quantum
  cryptography: {W}hat advancements in quantum computing mean for it
  professionals,'' \emph{IT Professional}, vol.~18, no.~5, pp. 42--47, 2016.

\bibitem{2}
F.~Xu, X.~Ma, Q.~Zhang, H.-K. Lo, and J.-W. Pan, ``Secure quantum key
  distribution with realistic devices,'' \emph{Reviews of Modern Physics},
  vol.~92, no.~2, p. 025002, 2020.

\bibitem{3}
C.~M. Gussen, P.~S. Diniz, M.~Campos, W.~A. Martins, F.~M. Costa, and J.~N.
  Gois, ``A survey of underwater wireless communication technologies,''
  \emph{J. Commun. Inf. Sys}, vol.~31, no.~1, pp. 242--255, 2016.

\bibitem{4}
S.-C. Zhao, X.-H. Han, Y.~Xiao, Y.~Shen, Y.-J. Gu, and W.-D. Li, ``Performance
  of underwater quantum key distribution with polarization encoding,''
  \emph{JOSA A}, vol.~36, no.~5, pp. 883--892, 2019.

\bibitem{5}
M.~Lanzagorta and J.~Uhlmann, ``Assessing feasibility of secure quantum
  communications involving underwater assets,'' \emph{IEEE Journal of Oceanic
  Engineering}, vol.~45, no.~3, pp. 1138--1147, 2019.

\bibitem{6}
A.~H.~F. Raouf, M.~Safari, and M.~Uysal, ``Performance analysis of quantum key
  distribution in underwater turbulence channels,'' \emph{JOSA B}, vol.~37,
  no.~2, pp. 564--573, 2020.

\bibitem{7}
P.~Shi, S.-C. Zhao, Y.-J. Gu, and W.-D. Li, ``Channel analysis for single
  photon underwater free space quantum key distribution,'' \emph{JOSA A},
  vol.~32, no.~3, pp. 349--356, 2015.

\bibitem{8}
M.~Lopes and N.~Sarwade, ``Optimized decoy state {QKD} for underwater free
  space communication,'' \emph{International Journal of Quantum Information},
  vol.~16, no.~02, p. 1850019, 2018.

\bibitem{9}
A.~H.~F. Raouf, M.~Safari, and M.~Uysal, ``Multi-hop quantum key distribution
  with passive relays over underwater turbulence channels,'' \emph{JOSA B},
  vol.~37, no.~12, pp. 3614--3621, 2020.

\bibitem{10}
S.~Zhao, W.~Li, Y.~Shen, Y.~Yu, X.~Han, H.~Zeng, M.~Cai, T.~Qian, S.~Wang,
  Z.~Wang \emph{et~al.}, ``Experimental investigation of quantum key
  distribution over a water channel,'' \emph{Applied optics}, vol.~58, no.~14,
  pp. 3902--3907, 2019.

\bibitem{11}
C.-Q. Hu, Z.-Q. Yan, J.~Gao, Z.-Q. Jiao, Z.-M. Li, W.-G. Shen, Y.~Chen, R.-J.
  Ren, L.-F. Qiao, A.-L. Yang \emph{et~al.}, ``Transmission of photonic
  polarization states through 55-m water: {T}owards air-to-sea quantum
  communication,'' \emph{Photonics Research}, vol.~7, no.~8, pp. A40--A44,
  2019.

\bibitem{12}
C.-Q. Hu, Z.-Q. Yan, J.~Gao, Z.-M. Li, H.~Zhou, J.-P. Dou, and X.-M. Jin,
  ``Decoy-state quantum key distribution over a long-distance high-loss
  air-water channel,'' \emph{Physical Review Applied}, vol.~15, no.~2, p.
  024060, 2021.

\bibitem{13}
Z.~Feng, S.~Li, and Z.~Xu, ``Experimental underwater quantum key
  distribution,'' \emph{Optics Express}, vol.~29, no.~6, pp. 8725--8736, 2021.

\bibitem{14}
C.~H. Bennett and G.~Brassard, ``Quantum cryptography: {P}ublic key
  distribution and coin tossing,'' in \emph{IEEE International Conference on
  Computers, Systems, and Signal Processing (IEEE, 1984)}, 1984, pp. 175--179.

\bibitem{15}
J.~Zhang, M.~A. Itzler, H.~Zbinden, and J.-W. Pan, ``Advances in {InGaAs/InP}
  single-photon detector systems for quantum communication,'' \emph{Light:
  Science \& Applications}, vol.~4, no.~5, pp. e286--e286, 2015.

\bibitem{16}
J.~Zhang, P.~Eraerds, N.~Walenta, C.~Barreiro, R.~Thew, and H.~Zbinden, ``2.23
  {GH}z gating {InGaAs/InP} single-photon avalanche diode for quantum key
  distribution,'' in \emph{Advanced Photon Counting Techniques IV}, vol.
  7681.\hskip 1em plus 0.5em minus 0.4em\relax International Society for Optics
  and Photonics, 2010, p. 76810Z.

\bibitem{17}
M.~Er-long, H.~Zheng-fu, G.~Shun-sheng, Z.~Tao, D.~Da-Sheng, and G.~Guang-Can,
  ``Background noise of satellite-to-ground quantum key distribution,''
  \emph{New Journal of Physics}, vol.~7, no.~1, p. 215, 2005.

\bibitem{18}
B.~Baek, L.~Ma, A.~Mink, X.~Tang, and S.~W. Nam, ``Detector performance in
  long-distance quantum key distribution using superconducting nanowire
  single-photon detectors,'' in \emph{Advanced Photon Counting Techniques III},
  vol. 7320.\hskip 1em plus 0.5em minus 0.4em\relax International Society for
  Optics and Photonics, 2009, p. 73200D.

\bibitem{19}
S.~Jiang, W.~O. Popoola, and M.~Safari, ``Quantum key distribution using
  time-gated {SPAD}s over turbid underwater channels,'' in \emph{CLEO: Science
  and Innovations}.\hskip 1em plus 0.5em minus 0.4em\relax Optical Society of
  America, 2021, pp. JW1A--125.

\bibitem{20}
R.~A. Leathers, T.~V. Downes, C.~O. Davis, and C.~D. Mobley, ``Monte {C}arlo
  radiative transfer simulations for ocean optics: {A} practical guide,'' Naval
  Research Lab Washington Dc Applied Optics Branch, Tech. Rep., 2004.

\bibitem{21}
W.~C. Cox~Jr, \emph{Simulation, modeling, and design of underwater optical
  communication systems}.\hskip 1em plus 0.5em minus 0.4em\relax North Carolina
  State University, 2012.

\bibitem{22}
M.~Lanzagorta, \emph{Underwater {C}ommunications}.\hskip 1em plus 0.5em minus
  0.4em\relax Morgan \& Claypool Publishers, 2012.

\bibitem{23}
H.~V. Nguyen, P.~V. Trinh, A.~T. Pham, Z.~Babar, D.~Alanis, P.~Botsinis,
  D.~Chandra, S.~X. Ng, and L.~Hanzo, ``Network coding aided cooperative
  quantum key distribution over free-space optical channels,'' \emph{IEEE
  Access}, vol.~5, pp. 12\,301--12\,317, 2017.

\bibitem{24}
Y.~Chen, S.~Huang, and M.~Safari, ``Orbital angular momentum multiplexing for
  free-space quantum key distribution impaired by turbulence,'' in \emph{Proc.
  IEEE Int. Wireless Commun. Mobile Comput. Conf.}\hskip 1em plus 0.5em minus
  0.4em\relax IEEE, 2018, pp. 636--641.

\bibitem{25}
C.~Gabriel, M.-A. Khalighi, S.~Bourennane, P.~L{\'e}on, and V.~Rigaud,
  ``Monte-{C}arlo-based channel characterization for underwater optical
  communication systems,'' \emph{Journal of Optical Communications and
  Networking}, vol.~5, no.~1, pp. 1--12, 2013.

\bibitem{26new}
F.~Hufnagel, A.~Sit, F.~Grenapin, F.~Bouchard, K.~Heshami, D.~England,
  Y.~Zhang, B.~J. Sussman, R.~W. Boyd, G.~Leuchs \emph{et~al.},
  ``Characterization of an underwater channel for quantum communications in the
  {O}ttawa river,'' \emph{Optics express}, vol.~27, no.~19, pp.
  26\,346--26\,354, 2019.

\bibitem{26}
V.~I. Haltrin, ``Two-term {H}enyey-{G}reenstein light scattering phase function
  for seawater,'' in \emph{IEEE International Geoscience and Remote Sensing
  Symposium. IGARSS'99 (Cat. No. 99CH36293)}, vol.~2.\hskip 1em plus 0.5em
  minus 0.4em\relax IEEE, 1999, pp. 1423--1425.

\bibitem{28}
C.~D. Mobley, \emph{Light and {W}ater: {R}adiative {T}ransfer in {N}atural
  {W}aters}.\hskip 1em plus 0.5em minus 0.4em\relax Academic press, 1994.

\bibitem{27}
C.~Mobley, E.~Boss, and C.~Roesler, ``Ocean optics web book,''
  \url{http://www.oceanopticsbook.info}, 2010.

\end{thebibliography}

\end{document}